\documentclass[twocolumn, twocolappendix]{openjournal}
\usepackage{xcolor}
\usepackage{xspace}
\usepackage{amsmath}
\usepackage{CJKutf8}
\usepackage{tabularx}
\usepackage{makecell}
\usepackage{multirow}
\usepackage{orcidlink}
\usepackage{natbib}
\usepackage{enumerate}
\usepackage{underscore}
\usepackage{amssymb}
\usepackage{graphicx}
\usepackage{enumitem}
\graphicspath{{Figures/}}
\usepackage{hyperref}

\begin{document}
\newcommand{\At}[1]{\left.#1\right|}
\newcommand{\sqb}[1]{\left[#1\right]}
\newcommand{\pr}[1]{\left(#1\right)}

\newcommand{\EQ}[1]{\textcolor{red}{(EQ: #1)}}
\newcommand{\EY}[1]{\textcolor{orange}{#1}}

\begin{CJK*}{UTF8}{gbsn}
\title{Steady-state Stellar Winds Driven by Recombination}
\author{Eritas Yang (杨晴) \orcidlink{0009-0005-2641-1531}}
\affiliation{Department of Astrophysical Sciences, Princeton University, 4 Ivy Lane, Princeton, NJ 08540, USA}
\email{eritas.yang@princeton.edu}

\author{Eliot Quataert (奎寅略) \orcidlink{0000-0001-9185-5044}}
\affiliation{Department of Astrophysical Sciences, Princeton University, 4 Ivy Lane, Princeton, NJ 08540, USA}
\email{quataert@princeton.edu}

\begin{abstract}
Hydrogen and helium recombination energy has been proposed as a potential driver of mass ejection in common-envelope evolution and other eruptive stellar phenomena. We investigate whether recombination can by itself launch a steady, transonic wind from near a stellar surface. Using a tabulated equation of state, we explore steady-state, adiabatic wind solutions over a broad range of stellar mass, density, and temperature. We classify a wind as recombination-driven only if the gas is gravitationally bound prior to recombination and if the released energy remains trapped until the flow becomes unbound. Only a small subset of the transonic solutions satisfies both conditions. In most cases, the gas is either already unbound without recombination or loses the released energy through radiative diffusion while still bound.
The remaining candidate solutions require substantial outflow velocities of order $\gtrsim10\,{\rm km\,s^{-1}}$ at $10\,R_\odot$ and mass-loss rates $\sim 0.1$-$10\,\rm M_\odot/yr$, which would rapidly deplete the stellar envelope.
We conclude that, within our adiabatic framework and subject to an a posteriori radiative-trapping criterion, recombination energy alone is unlikely to launch a steady wind from a hydrostatic envelope. It can, however, accelerate and unbind a pre-existing outflow generated by processes such as binary orbital decay.
\end{abstract}

\section{Introduction}

Recombination energy stored in partially ionized gas has long been recognized as a potentially important energy reservoir in various stellar contexts.  In the late 1960s, Paczy{\'n}ski found that the envelope binding energy of certain red supergiants becomes positive when hydrogen and helium recombination energy is included, and speculated that recombination could help unbind convective envelopes via a wind \citep{Paczynski1968, Paczynski1969}.
More recently, the role of recombination has been reconsidered in the context of common-envelope evolution, as hydrodynamical simulations demonstrate that the orbital energy released by the in-spiraling companion is often insufficient to unbind the envelope on a dynamical timescale \citep[e.g.,][]{Passy_2012, Ricker_2012, Kuruwita_2016, Staff_2016, Ohlmann_2016, Iaconi_2018}.

The extent to which recombination energy assists envelope ejection remains debated. Some studies argue that it can make a significant contribution, especially when released in optically thick regions where the energy remains trapped \citep[e.g.,][]{Nandez_2016,Ivanova_2018,Sand_2020,Lau_2022}.
Others argue that a substantial fraction of this energy is radiated away and therefore does not directly contribute to mechanical ejection \citep[e.g.,][]{Sabach_2017, Grichener_2018, Soker_2018, Lau_2025}, but instead primarily regulates radiation transport and the thermal structure of the expanding gas. 

Despite this extensive literature, it remains unclear whether recombination energy can launch an outflow by itself or instead primarily modifies gas that has already begun expanding. In many common-envelope simulations, recombination occurs only after the envelope has entered dynamical expansion, making it difficult to distinguish these possibilities. The same question arises in eruptive mass loss from evolved stars, where recombination can contribute mechanically only if the released energy remains trapped while the gas is still gravitationally bound.

Here we investigate whether recombination can drive a steady, transonic outflow from near a stellar surface. Although common-envelope evolution and eruptive mass loss are inherently time-dependent, the steady-state framework allows us to isolate the energetic and radiative requirements for wind launching and to explore them systematically over a broad parameter space. We find that recombination alone rarely produces a viable wind: the gas is generally either already unbound before recombination or able to radiate the released energy before becoming unbound. 
The remaining candidate solutions already start with a substantial coherent outflow at $10\,R_\odot$. Recombination therefore does not provide the mechanism that initiates these flows from hydrostatic equilibrium; instead, it further accelerates and unbinds gas that has already been set in outward motion by another process, such as energy deposition from binary orbital decay in a common-envelope event or stellar merger.

In Section~\ref{sec:H_and_He}, we describe our wind model and define the criteria we impose for a recombination-driven wind. We present the parameter survey and resulting wind solutions in Section~\ref{sec:sol}. We then discuss the implications and caveats of our results in Section~\ref{sec:disc}.

\section{Recombination-driven wind model} \label{sec:H_and_He}
\subsection{Adiabatic wind model} \label{ssec:model}
We assume that recombination proceeds adiabatically, such that the released energy is retained as internal energy of the gas. This assumption provides a ``best-case'' limit for wind driving: any radiative loss will reduce the thermal and mechanical energy available to accelerate the flow.
Under the adiabatic assumption, the temperature and density of the system evolve as
\begin{equation}
    \frac{dT}{T} = (\gamma_3-1)\frac{d\rho}{\rho}, \label{eq:EoS}
\end{equation}
where $\gamma_3$ is the third generalized adiabatic index. In this formulation, recombination energy is included self-consistently in the equation of state (EoS), modifying both the internal energy and the adiabatic indices. Recombination can in principle contribute to a wind by driving $\gamma_3 \rightarrow 1$ over a range of temperatures, effectively creating thermodynamical conditions analogous to those of an isothermal wind. The tabulated EoS used for our numerical calculations is described in Section~\ref{sec:sol}, and explicit analytical expressions for a hydrogen-only EoS are given in Section~\ref{ssec:H}.

We now consider a steady, spherically symmetric wind. The momentum equation is
\begin{equation}
    v\frac{dv}{dr} + \frac{1}{\rho}\frac{dp}{dr} + \frac{GM_r}{r^2} = 0,
    \label{eq:momentum}
\end{equation}
where $p=p_{\rm rad} + p_{\rm gas}$ is the total pressure\footnote{The radiation-pressure gradient is included, but direct acceleration by the diffusive radiative flux, $\kappa_R F_{\rm diff}/c$, is neglected. In the adiabatic limit considered here, radiation is trapped and contributes through pressure and enthalpy rather than diffusive flux driving (see eq.~\ref{eq:tratio}).}, $M_r$ is the enclosed mass within radius $r$, and
\begin{equation}
    \frac{dM_r}{dr} = 4\pi r^2\rho.
\end{equation}
We explicitly include the self-gravity of the gas, because the integrated gas mass in some wind solutions becomes non-negligible.

Under mass conservation, $\dot{M}=4\pi r^2 \rho v = \rm const.$, the pressure gradient can be written as
\begin{equation}
    \frac{dp}{dr} = \frac{dp}{d\rho}\cdot \frac{d\rho}{dr} = -c_s^2 \left(\frac{2\rho}{r} + \frac{\rho}{v}\frac{dv}{dr}\right),
    \label{eq:dp}
\end{equation}
where $c_s^2 \equiv \At{\frac{dp}{d\rho}}_S$ is the adiabatic sound speed.
We can then rewrite Equations~\eqref{eq:EoS} and \eqref{eq:momentum} into a first-order system of differential equations:
\begin{equation}
    \frac{d\ln T}{dr} = -(\gamma_3-1) \pr{\frac{2}{r} + \frac{d\ln v}{dr}},
    \label{eq:dTdr}
\end{equation}
\begin{equation}
    \frac{d\ln v}{dr} = \left(\frac{2c_s^2}{r}-\frac{GM_r}{r^2}\right)/(v^2-c_s^2).
    \label{eq:dvdr}
\end{equation}

Equation~\eqref{eq:dvdr} has a formal singularity at $v=c_s$, since the velocity gradient diverges unless the numerator simultaneously vanishes. Requiring the numerator and denominator to vanish together therefore defines the critical condition for a smooth transonic solution:
\begin{equation}
    c_s^2 = v^2 = \frac{GM_r}{2r} \quad \text{at the critical point.} \label{eq:crit_condition}
\end{equation}
For a given choice of density, temperature, and enclosed mass at the critical point, $\{\rho_c, T_c, M_{r,c}\}$, the critical radius and velocity are
\begin{equation}
    r_c = \frac{GM_{r,c}}{2 c_s^2(\rho_c, T_c)}, \quad v_c=c_s(\rho_c,T_c), \label{eq:critical}
\end{equation}
where the sound speed $c_s$ is determined by the EoS.
The velocity gradient at the critical point is then determined by applying L'H\^{o}pital's rule to Equation~\eqref{eq:dvdr}, and a transonic wind solution has  $(dv/dr)_{r_c} > 0$.

\subsection{Validity criteria \label{ssec:criteria}}
We impose two additional conditions for a transonic solution to qualify as a recombination-driven wind. First, our wind model assumes that recombination energy is fully retained as internal energy of the gas rather than lost radiatively. This requires that the thermal diffusion timescale exceed the advection timescale, $t_{\rm th} > t_{\rm adv}$.
Equivalently, the advected energy flux must exceed the diffusive radiative flux,
\begin{equation}
    \frac{t_{\rm th}}{t_{\rm adv}} \approx 
    \frac{F_{\rm adv}}{F_{\rm diff}} 
    =
    \frac{\rho v \cdot h}
    {\frac{c}{\rho\kappa_R} \left|\frac{dp_{\rm rad}}{dr}\right|} > 1,
    \label{eq:tratio}
\end{equation}
where $h=u+p/\rho$ is the total specific enthalpy, with $u$ being the specific internal energy including the latent heat of recombination, and $\kappa_R$ is the Rosseland mean opacity.
Here we adopt the pre-computed opacity tables for solar composition from {\AE}SOPUS \citep{Marigo_2009} to estimate the importance of radiative losses a posteriori in our wind solutions that do not include radiative diffusion.

For an adiabatic recombination-driven outflow to be  physical, we require $F_{\rm adv} \gtrsim F_{\rm diff}$ until the flow reaches the local escape velocity ($v \approx v_{\rm esc}$).
If $F_{\rm adv}/F_{\rm diff} < 1$ while the gas is still gravitationally bound, the energy required to unbind the envelope is lost to radiation. To order unity, this condition is equivalent to requiring that the optical depth satisfy $\tau \gtrsim (c/v)(u_{\rm rad}/u_{\rm tot})$ in the region where recombination accelerates the flow to above the escape speed.

Second, we require that the energy from recombination be dynamically important in driving the outflow. The total energy budget per unit mass available to the gas is described by the Bernoulli constant,
\begin{equation}
    \mathcal{B} \equiv \frac{v^2}{2} + \Phi + h,
    \label{eq:Be}
\end{equation}
where $\Phi$ is the gravitational potential energy, $\Phi(r) = -\int_r^\infty \frac{GM_{r'}}{r'^2}dr'$.
For a steady, adiabatic flow, $\mathcal{B}$ is conserved along streamlines.\footnote{More generally, when radiative diffusion is included, conservation of the total energy flux gives $\dot{E} = \dot{M}\mathcal{B} + L_{\rm diff} = {\rm const}.$ The energy required to lift and accelerate the gas can therefore be supplied either by the advected enthalpy within $\dot{M}\mathcal{B}$, or by the diffusive radiative luminosity. In the adiabatic limit considered here, $L_{\rm diff}$ is not included and the wind is powered by enthalpy generated by recombination energy.}

The enthalpy contains contributions from the gas, radiation, and the latent energy stored in ionization,
\begin{equation}
    h = h_{\rm gas} + h_{\rm rad} + h_{\rm rec}.
\end{equation}
As the gas recombines, $h_{\rm rec}$ decreases and the released energy can be used to accelerate the gas.
To isolate its contribution, we define a modified Bernoulli parameter $\tilde{\mathcal{B}}$ by explicitly removing the recombination energy term,
\begin{equation}
    \tilde{\mathcal{B}} \equiv \mathcal{B} - h_{\rm rec} = \mathcal{B} - \frac{1}{\rho}\sum_i n_i \epsilon_i,
    \label{eq:Betilde}
\end{equation}
with $\epsilon_i =$ 13.6, 24.6 and 54.4~eV corresponding to the
recombination energies of
${\rm H^+}\rightarrow{\rm H^0}$,
${\rm He^+}\rightarrow{\rm He^0}$, and
${\rm He^{2+}}\rightarrow{\rm He^+}$, respectively.

The ratio and sign of $\tilde{\mathcal{B}}/\mathcal{B}$ therefore provide a direct measure of the contribution of recombination energy to the overall energy budget of the flow.
In particular, solutions starting with $\tilde{\mathcal{B}}>0$ are energetically unbound even in the absence of recombination. Conversely, winds with $\tilde{\mathcal{B}}<0$ prior to recombination must rely on the release of recombination energy in order to become unbound.

Our validity criteria for recombination-driven wind solutions can thus be summarized as follows:
\begin{enumerate}[label=(\roman*)]
    \item $F_{\rm adv} > F_{\rm diff}$ at $v = v_{\rm esc}$.
    \item $\tilde{\mathcal{B}}<0$ in the fully ionized region.
\end{enumerate}

\subsection{Analytic estimates}\label{ssec:estimates}
Before presenting the numerical solutions, we derive several estimates that delimit the relevant parameter space.
In order for the flow to pass through the critical point and become supersonic, the velocity gradient must satisfy $dv/dr >0$ at $r=r_c$. As shown in Appendix~\ref{app:dvdr}, this requires
\begin{equation}
    -2 \lesssim A_c \lesssim \frac{1}{2} \label{eq:A_condition}
\end{equation}
at the critical point, where
\begin{equation}
    A_c \equiv \sqb{ \At{\frac{d\ln c_s^2}{d\ln\rho}}_T + (\gamma_3-1) \At{\frac{d\ln c_s^2}{d\ln T}}_\rho}_{r_c}.  \label{eq:A}
\end{equation}
For a simple polytropic EoS, $P \propto \rho^\gamma$, the sound speed scales as $c_s^2 \propto \rho^{\gamma-1}$, giving $A_c = \gamma - 1$. The critical-point condition in Equation~\eqref{eq:A_condition} then translates to $\gamma < 3/2$.  One physical interpretation is that in a constant-speed wind, $\rho \propto r^{-2}$, so the thermal energy declines as $\rho^{\gamma-1}\propto r^{-2(\gamma-1)}$. Thus, $\gamma = 3/2$ is the critical value below which the thermal energy decreases more slowly than the gravitational potential energy, allowing an accelerating solution at the critical point.

A complementary constraint comes from the finite amount of recombination energy available to power the outflow. If this energy is fully converted into kinetic energy, we may define a characteristic wind velocity, $\frac{1}{2} v_{\epsilon}^2 = h_{\rm rec}$, where $h_{\rm rec}$ is evaluated for fully ionized gas, i.e., prior to recombination. For solar composition, this gives $v_\epsilon \approx 54\,{\rm km\,s^{-1}}$. This velocity provides an approximate upper limit on the terminal speed achievable by a recombination-powered wind. 
Therefore, only stars or merger remnants with escape velocities $v_{\rm esc} \lesssim v_{\epsilon}$ near the hydrogen or helium recombination zones can plausibly launch outflows powered primarily by recombination energy. This corresponds to giants with recombination occurring at
\begin{equation}
    r_{\rm rec} \gtrsim \frac{2GM_r}{v_\epsilon^2} \simeq 130\,R_\odot \pr{\frac{M_r}{M_\odot}}\pr{\frac{v_\epsilon}{54\,\rm km\,s^{-1}}}^{-2}.
\end{equation}

We can also estimate the mass-loss rate of the wind by evaluating it at the critical point:
\begin{equation}
    \dot{M} = \At{4\pi r^2\rho v }_{r_c} \approx \At{\frac{ 4\pi\tau}{\kappa} \frac{GM_r}{2c_s}}_{r_c},
\end{equation}
where we have used the critical-point condition $v^2=c_s^2=GM_r/2r$ together with $\tau \sim \kappa \rho r$. Expressing this in terms of Eddington luminosity\footnote{The appearance of $L_{\rm Edd}$ does not imply a comparable diffusive luminosity. As mentioned in Section~\ref{ssec:criteria}, in the trapped regime, the wind is powered primarily by conversion of advected enthalpy.},
\begin{equation}
\begin{aligned}
    \dot{M} & \approx \At{\frac{L_{\rm edd}}{2c_s^2} \frac{\tau}{c/v}}_{r_c} \sim 0.3\,{\rm M_\odot\,yr^{-1}} \times \\
    & \frac{M/M_\odot}{\sqb{\kappa/\kappa_T} \sqb{c_s/20\,{\rm km\,s^{-1}}}^2} \At{\frac{\tau}{c/v}}_{r_c}
\end{aligned} \label{eq:Mdotestimate}
\end{equation}
where $\kappa_T$ is the electron scattering opacity, and the adopted sound speed corresponds to temperatures characteristic of the region between hydrogen and helium recombination.\footnote{Electron scattering provides a reasonable opacity estimate in helium recombination zones, though the opacity is generally smaller during hydrogen recombination.}
The adiabaticity requirement implies $\frac{\tau}{c/v} \gtrsim u_{\rm rad}/u_{\rm tot}$ in the recombination region near the critical point. Because the numerical solutions presented in the next section typically have $u_{\rm rad} \sim u_{\rm gas}$, Equation~\eqref{eq:Mdotestimate} implies characteristic mass-loss rates of order ${\rm M_\odot\,yr^{-1}}$. Any steady phase at such a rate must be short-lived relative to the finite mass reservoir of a stellar envelope.


\section{Wind solutions}\label{sec:sol}

We next numerically explore wind solutions using the model developed in Section~\ref{ssec:model}.
We adopt the tabulated EoS from \citet{Andalman_2025}, which self-consistently includes hydrogen and helium recombination, H$_2$ formation, and radiation pressure, assuming a helium mass fraction of $Y=0.3$. The EoS is tabulated in density and temperature, which we take to be the independent thermodynamic variables. All other quantities required by the wind equations -- including pressure, internal energy, ionization fractions, and thermodynamic derivatives -- are obtained by interpolation within the table.

To explore the parameter space for candidate wind solutions, we sample critical-point conditions (see eq.~\ref{eq:critical}) over the ranges $\rho_c\in[10^{-15},10^{-5}]\,{\rm g\,cm^{-3}}$, $T_c\in[10^3,10^6]\,{\rm K}$, and $M_{r,c}\in[1,50]\,M_\odot$. 
Our calculation is not matched to a particular stellar structure model. Instead, it identifies the local thermodynamic and gravitational conditions under which recombination-assisted transonic solutions can exist. Whether these conditions are realized within a self-consistent stellar envelope must therefore be assessed separately. We adopt this more general approach because our primary applications are common-envelope events and stellar mergers, where the envelope can be driven far from hydrostatic equilibrium and the range of structures produced during the interaction is not yet well characterized.

\begin{figure}
    \centering
    \includegraphics[width=0.8\linewidth]{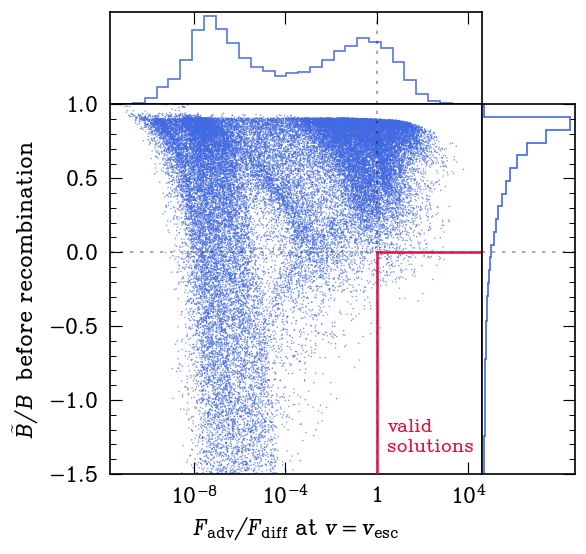}
    \caption{The full ensemble of the mathematically valid transonic solutions.
    The horizontal axis shows $F_{\rm adv}/F_{\rm diff}$ evaluated at $v=v_{\rm esc}$, and the vertical axis shows $\tilde{\mathcal{B}}/\mathcal{B}$ evaluated prior to recombination.
    Blue points show all 46,000 solutions that integrate successfully across the full radial domain. The red box marks the 142 candidate recombination-driven winds that satisfy both criteria in Section~\ref{ssec:criteria}: radiative trapping ($F_{\rm adv}/F_{\rm diff}>1$ at $v=v_{\rm esc}$) and energetically bound gas prior to recombination ($\tilde{\mathcal{B}}<0$).
    The full ensemble is shown to illustrate why most transonic solutions do not correspond to recombination-driven winds.}
    \label{fig:valid_region}
\end{figure}

For each set of critical-point conditions, we integrate Equations~\eqref{eq:dTdr} and \eqref{eq:dvdr} inward from $r=r_c$ to $10\,R_\odot$ and outward to $10^5\,R_\odot$. Of the 46,000 solutions that integrate successfully across this radial domain, only 142 ($\simeq 0.3\%$) satisfy both criteria in Section~\ref{ssec:criteria} and thus qualify as candidate recombination-driven winds. Figure~\ref{fig:valid_region} shows the distribution of solutions in the parameter space defined by these two criteria.

\begin{figure*}
    \centering
    \includegraphics[width=1.0\linewidth]{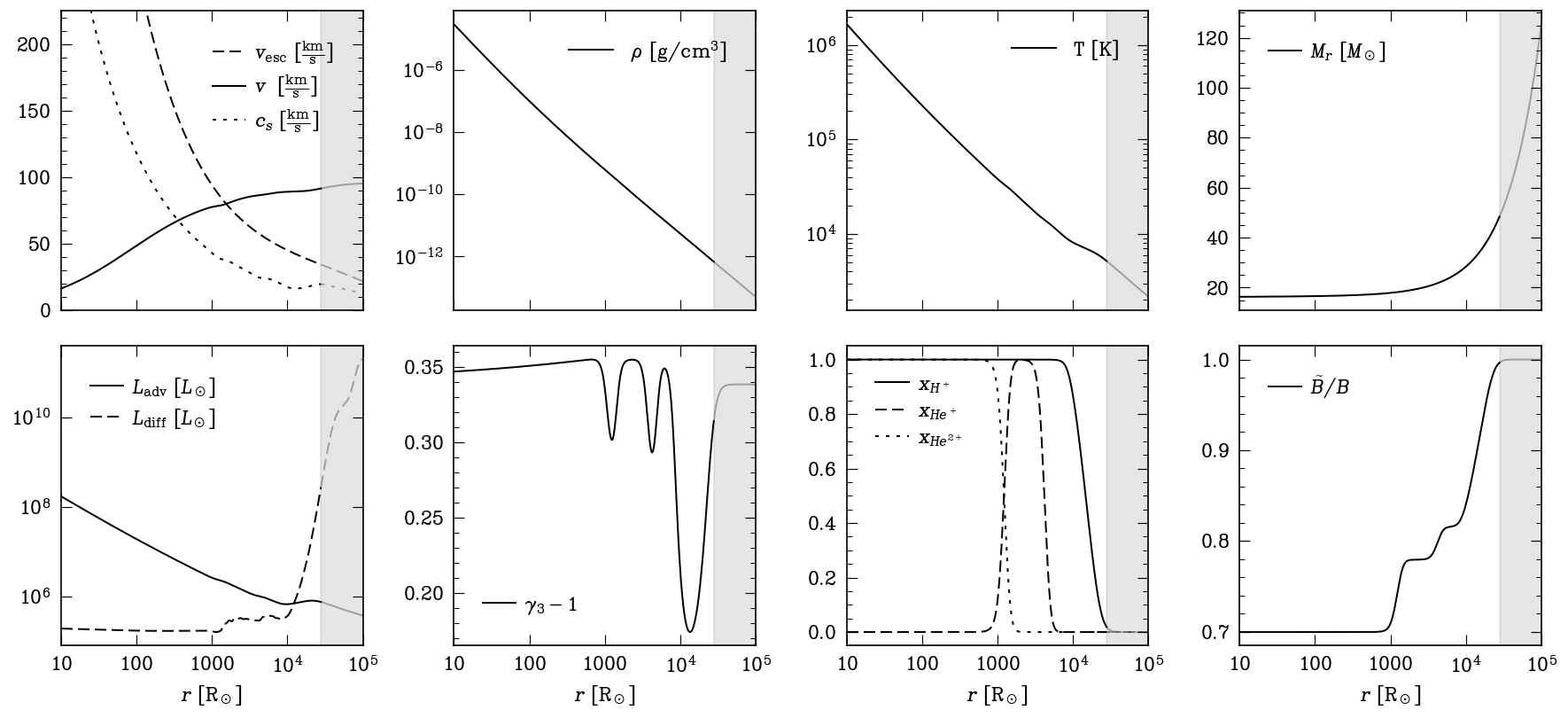}
    \caption{Example wind solution for a $13\,M_\odot$ star with a mass-loss rate of $\dot{M}=4.5\,M_\odot\,{\rm yr^{-1}}$. 
    The gray-shaded regions lie exterior to the photosphere, where radiation begins to free stream and our model is no longer applicable.
    \textbf{Top:} radial profiles of the wind velocity $v$ compared with the escape velocity $v_{\rm esc}$ and sound speed $c_s$, the gas density $\rho$, the temperature $T$, and the enclosed mass $M_r$. \textbf{Bottom:} radial profiles of the advective and diffusive luminosities, the effective adiabatic index $\gamma_3-1$, the hydrogen and helium ionization fractions $x_i$, and the scaled Bernoulli parameter $\tilde{\mathcal{B}}/\mathcal{B}$ (eqs.~\ref{eq:Be} and \ref{eq:Betilde}). Because $\tilde{\mathcal{B}}>0$ prior to recombination, the gas is already energetically unbound without recombination energy. Note that the wind profiles are obtained from the strictly adiabatic model, while $L_{\rm diff}$ is a post-processed estimate used only to assess radiative losses and when they modify the wind solution. Once $L_{\rm diff}\gtrsim L_{\rm adv}$, the adiabatic temperature profile is no longer self-consistent, so larger values of $L_{\rm diff}$ are not physical.}
    \label{fig:adiabatic_case}
\end{figure*}

\begin{figure*}
    \centering
    \includegraphics[width=1.0\linewidth]{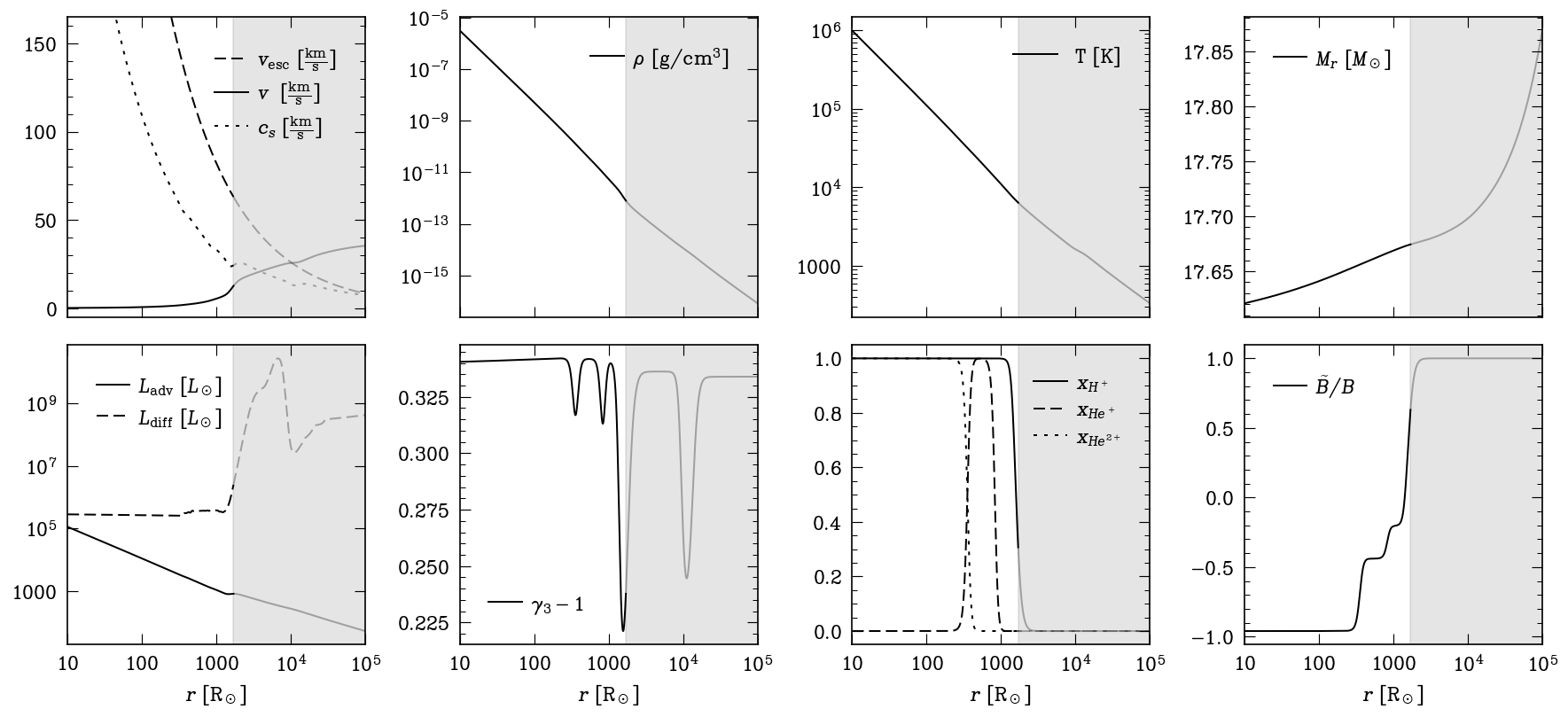}
    \caption{Same as Figure~\ref{fig:adiabatic_case}, but for a $17\,M_\odot$ star with a mass-loss rate of $\dot{M}=0.003\,M_\odot\,{\rm yr^{-1}}$. The flow is energetically bound prior to recombination (bottom right), but radiative diffusion becomes efficient well before the outflow reaches the escape velocity (bottom left). The recombination energy is therefore expected to escape rather than be converted into wind kinetic energy.}
    \label{fig:bound_case}
\end{figure*}

Most solutions fail in one of two ways: either the gas is already energetically unbound without recombination, or radiative diffusion becomes important while the gas remains bound. Figure~\ref{fig:adiabatic_case} illustrates the first case. Radiative diffusion remains inefficient out to $r\sim10^4\,R_\odot$, but $\tilde{\mathcal{B}}>0$ prior to recombination, so the outflow does not require recombination energy to become unbound. Figure~\ref{fig:bound_case} illustrates the second case. Here the flow is initially bound, but radiative diffusion is important even near the inner boundary, allowing the recombination energy to escape before it can unbind the gas.

\begin{figure*}
    \centering
    \includegraphics[width=1.0\linewidth]{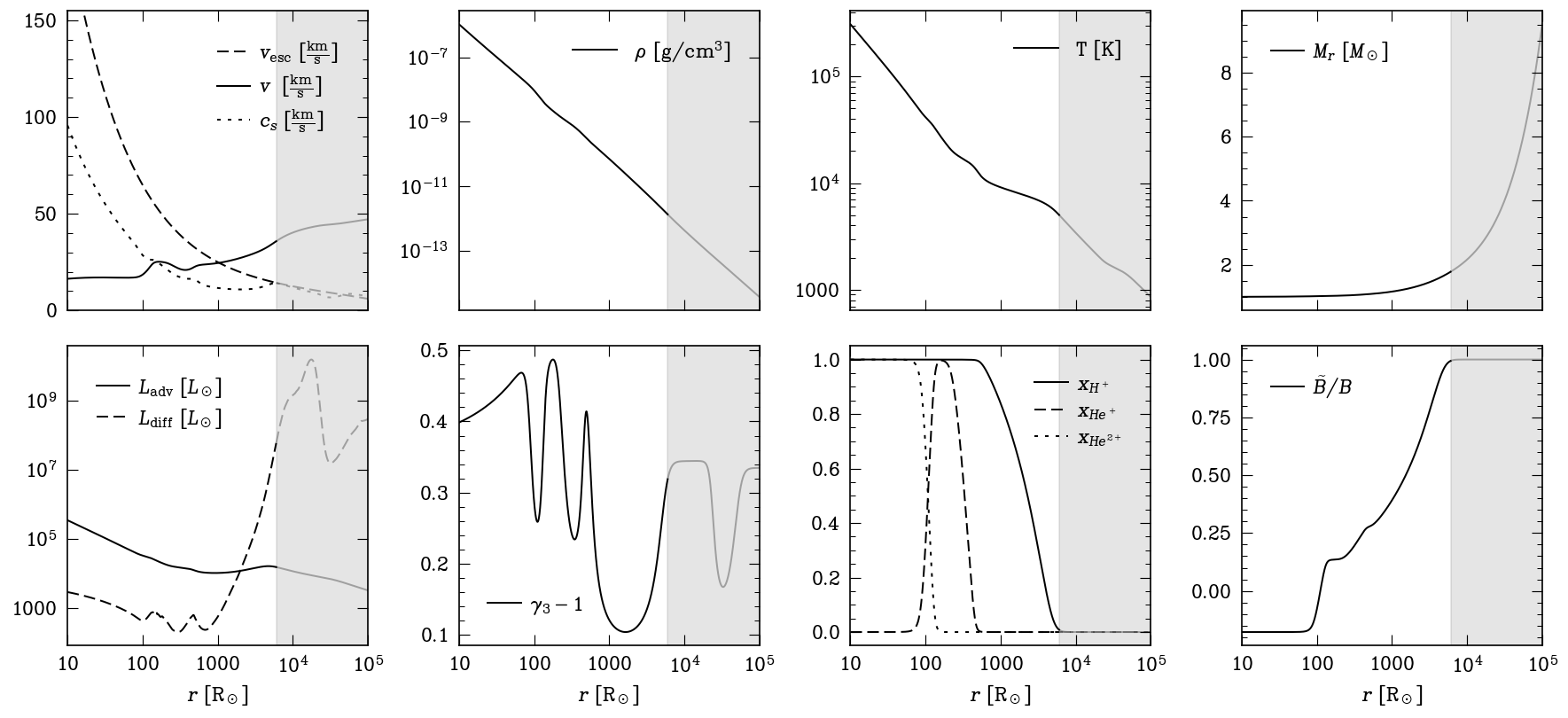}
    \caption{Same as Figure~\ref{fig:adiabatic_case}, but for a $1\,M_\odot$ star with a mass-loss rate of $\dot{M}=0.2\,M_\odot\,{\rm yr^{-1}}$. This solution satisfies both criteria in Section~\ref{ssec:criteria}, but requires an outflow velocity of $18\,\mathrm{km\,s^{-1}}$ at $r=10\,R_\odot$.}
    \label{fig:survived_case}
\end{figure*}

Figure~\ref{fig:survived_case} shows a representative candidate for a recombination-driven wind. The flow is slightly bound prior to recombination, and radiative diffusion does not become important until $r\sim2000\,R_\odot$, by which point recombination has accelerated the flow beyond the escape velocity. 
However, the solution already carries a substantial outward mass flux, with $\dot M=0.2\,\rm M_{\odot}\,yr^{-1}$ and $v=18\,{\rm km\,s^{-1}}$ at $r=10\,R_\odot$. Its density also approaches $\rho\propto r^{-2}$, as expected for a steady outflow with slowly varying velocity. Thus, the solution resembles an already-established outflow rather than a wind that emerges smoothly from a hydrostatic inner atmosphere. Recombination can accelerate and unbind this flow, but the solution does not demonstrate that recombination itself generates the underlying mass flux.

\subsection{Properties of candidate solutions}

\begin{figure*}
    \centering
    \includegraphics[width=1.0\linewidth]{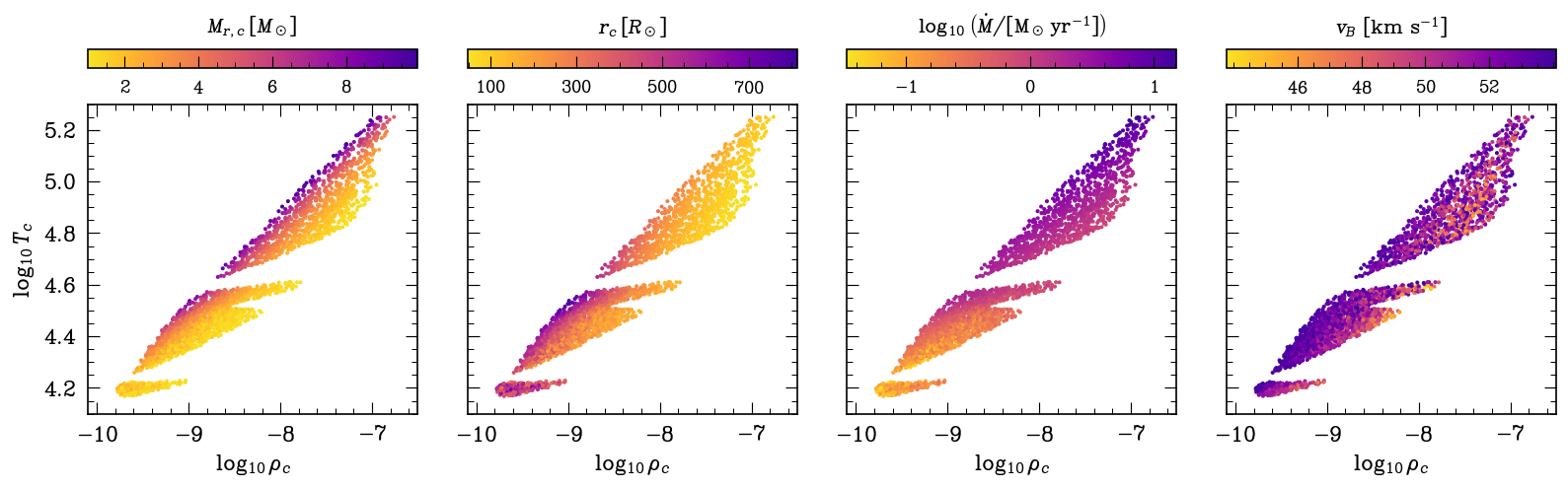}
    \caption{Properties of candidate recombination-driven winds after dense resampling of the region identified in Figure~\ref{fig:valid_region}. Each point satisfies both criteria in Section~\ref{ssec:criteria}. From left to right, the color scales show the enclosed mass at the critical point, the critical radius, the mass-loss rate, and the Bernoulli velocity $v_B\equiv\sqrt{2\mathcal{B}}$.}
    \label{fig:param_space}
\end{figure*}

We refine the sampling within the narrow region occupied by the candidate solutions in Figure~\ref{fig:valid_region}. The resulting solutions are shown in Figure~\ref{fig:param_space}. They span critical-point densities $\rho_c\sim10^{-10}$--$10^{-7}\,{\rm g\,cm^{-3}}$ and temperatures $T_c\sim10^{4.1}$--$10^{5.3}\,{\rm K}$, and occur only for enclosed masses $M_{r,c}\lesssim10\,M_\odot$. Most have critical radii $r_c\gtrsim100\,R_\odot$, placing the sonic point within the envelope only for the most extended red giants and AGB stars. Their mass-loss rates are also large, consistent with the estimate in Section~\ref{ssec:estimates}, so any steady wind phase would rapidly deplete the finite envelope mass.

\begin{figure}
    \centering
    \includegraphics[width=0.75\linewidth]{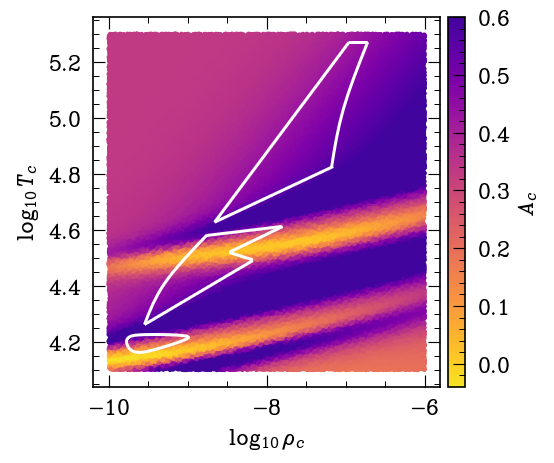}
    \caption{Map of $A_c$ (eq.~\ref{eq:A}) in the $\rho_c$--$T_c$ plane. White contours enclose the candidate solutions shown in Figure~\ref{fig:param_space}. The requirement $A_c\lesssim1/2$ for an accelerating transonic solution (eq.~\ref{eq:A_condition}) divides the allowed parameter space into three islands.}
    \label{fig:A}
\end{figure}

The candidate solutions separate into three islands in the $\rho_c$--$T_c$ plane. As illustrated in Figure~\ref{fig:A}, the intervening gaps occur where the critical-point condition $A_c\lesssim1/2$ is not satisfied (eq.~\ref{eq:A_condition}). These gaps therefore reflect changes in the thermodynamic derivatives of the EoS across the hydrogen and helium recombination regimes.

\begin{figure}
    \centering
    \includegraphics[width=0.8\linewidth]{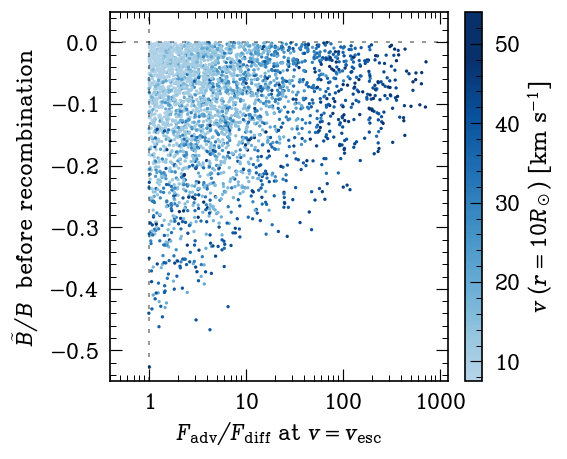}
    \caption{Candidate recombination-driven winds after dense resampling of the region in Figure~\ref{fig:valid_region}. The color scale shows the inner-boundary velocity at $r=10\,R_\odot$; all candidates have $v_0\gtrsim10\,\mathrm{km\,s^{-1}}$.}
    \label{fig:v_hist}
\end{figure}

Satisfying the two criteria in Section~\ref{ssec:criteria} is not sufficient to demonstrate that recombination can launch a wind from hydrostatic equilibrium. Figure~\ref{fig:v_hist} shows that every candidate solution already carries a coherent outward flow with $v_0\gtrsim10\,\mathrm{km\,s^{-1}}$ at $r=10\,R_\odot$. Combined with the mass-loss rates $\dot{M}\sim0.1$-$10\,\rm M_\odot/yr$, these solutions therefore require an already-established mass flux. Thus, coherent outward motion must already be established deep within the envelope before recombination can further accelerate and unbind the gas.

The candidate solutions also carry a large advected energy flux, $\dot M\mathcal{B}\sim10^4$--$10^6\,L_\odot$, typically exceeding the quiescent luminosities of the stars relevant here.
This does not by itself make the instantaneous wind solutions energetically inconsistent, since this energy is carried as enthalpy created by recombination.  However, the finite recombination energy reservoir cannot sustain the flow indefinitely. Together with the large mass-loss rates, this suggests that the candidate solutions should be interpreted as local or quasi-steady descriptions of short-lived eruptive outflows rather than long-lived stationary winds.

\subsection{Contribution from hydrogen vs. helium recombination} \label{ssec:H}
We can evaluate the relative importance of hydrogen and helium recombination with a hydrogen-only (i.e., $X=1$) wind model. Assuming local thermodynamic equilibrium (LTE), the total particle number density is $n=\frac{\rho}{m_p}(1+x)$, where the hydrogen ionization fraction $x\equiv {n(H^+)}/\sqb{n(H^0) + n(H^+)}$ is determined by the Saha equation:
\begin{equation}
    \frac{n(e^-)n(H^+)}{n(H^0)}=\frac{x^2}{1-x}\frac{\rho}{m_p}=\frac{(2\pi m_e kT)^{3/2}}{h^3}e^{-\epsilon_H/kT}, \label{eq:Saha}
\end{equation}
where $\epsilon_H = 13.6$~eV is the recombination energy.

Assuming that the released recombination energy is retained as internal energy of the gas, the total pressure and specific internal energy are
\begin{align}
    &p = (1+x)\frac{\rho}{m_p}kT + \frac{a_{\rm rad} T^4}{3}, \label{eq:p} \\
    &u = (1+x)\frac{3}{2}\frac{kT}{m_p} + \frac{a_{\rm rad}T^4}{\rho} + \frac{\epsilon_H x}{m_p}, \label{eq:u}
\end{align}
where $a_{\rm rad} \equiv 4\sigma/c$ is the radiation constant. In the adiabatic limit, $du =(p/\rho^2)d\rho$. Combining this condition with the logarithmic derivative of Equation~\eqref{eq:Saha} and with Equation~\eqref{eq:u} gives us the adiabatic index in Equation~\eqref{eq:EoS}:
\begin{equation}
    \gamma_3 = 1 + \frac{(1+x)(1+4\alpha) + \frac{x(1-x)}{(2-x)}(\frac{3}{2}+\frac{\epsilon_H}{kT})}{(1+x)(\frac{3}{2}+12\alpha) + \frac{x(1-x)}{(2-x)}(\frac{3}{2}+\frac{\epsilon_H}{kT})^2},
\end{equation}
where $\alpha = \frac{p_{\rm rad}}{p_{\rm gas}}$ (see also \citealp{Kasen_2010}).

The adiabatic sound speed is
\begin{equation}
    c_s^2 \equiv \At{\frac{dp}{d\rho}}_S
    = \At{\frac{dp}{d\rho}}_T + \At{\frac{dp}{dT}}_\rho \cdot \At{\frac{dT}{d\rho}}_S .
\end{equation}
Substituting in Equations~\eqref{eq:EoS} and \eqref{eq:p}, we find
\begin{equation}
\begin{aligned}
    &c_s^2(\rho, T)  =  (1+x)\frac{kT}{m_p}  \left[\gamma_3 - \frac{x(1-x)}{(2-x)(1+x)}\right. \\
     &\left.+ (\gamma_3-1)\frac{x(1-x)}{(2-x)(1+x)}\left(\frac{3}{2} + \frac{\epsilon_H}{kT}\right) + 4\alpha(\gamma_3-1)\right]. \label{eq:cs}
\end{aligned}
\end{equation}

We then apply these thermodynamic relations to the wind framework developed in Section~\ref{ssec:model} and integrate Equations~\eqref{eq:dTdr} and \eqref{eq:dvdr}. We sample critical-point conditions over the same parameter ranges explored in Section~\ref{sec:sol}. For each solution, we compute the optical depth using Rosseland mean opacity tables from the Opacity Project \citep{OP}, assuming a pure-hydrogen composition ($X=1$).

We find only 47 solutions that satisfy both validity criteria, compared to 142 when helium recombination is included\footnote{Because the EoS implementation and opacity treatment also differ, this comparison should be interpreted qualitatively.}, though in both cases these represent a vanishingly small fraction of the full ensemble of sampled solutions.
All 47 candidate solutions emerge with large inner-boundary velocities of $\approx 40\,\rm km\,s^{-1}$ at $r=10R_\odot$. Moreover, hydrogen recombination occurs well outside the sonic point, $r_{\rm recomb}\gg r_c$. Hydrogen recombination therefore does not launch these winds; instead, it increases the terminal velocity of an outflow that is already established.

This picture is consistent with studies of common-envelope evolution in which helium recombination, occurring deeper in the optically thick envelope, contributes more effectively to expansion, whereas much of the hydrogen recombination energy is released after the material is unbound or can escape from the outer envelope \citep[e.g.,][]{Soker_2018, Reichardt_2020, Lau_2022}.

\section{Discussion} \label{sec:disc}

In this work we show that hydrogen and helium recombination energy alone is unlikely to launch a steady stellar wind. Most mathematical transonic solutions fail at least one of the two requirements for recombination driving: either the gas is already energetically unbound when the latent ionization energy is excluded, or radiative diffusion becomes important before the gas reaches escape.
The candidate solutions that satisfy both criteria occupy a restricted region of thermodynamic parameter space, with critical-point conditions associated with hydrogen and helium recombination (Figs.~\ref{fig:valid_region} and \ref{fig:param_space}). 
These solutions nevertheless require large mass-loss rates $\dot M\sim0.1$--$10\,{\rm M_\odot\,yr^{-1}}$, which would rapidly deplete a stellar envelope, and they require outward velocities $\gtrsim10\,{\rm km\,s^{-1}}$ at $10\,R_\odot$ (Fig.~\ref{fig:v_hist}).

The distinction between \textit{launching} and \textit{accelerating} an outflow is central to interpreting these results. Our steady solutions do not provide a mechanism for generating the required mass flux from a hydrostatic envelope; instead, they show that once an outward flow has been established, recombination can provide additional enthalpy that accelerates and unbinds the gas. Convective velocities in the envelopes of red giants, AGB stars, and red supergiants can reach tens of ${\rm km\,s^{-1}}$ \citep[e.g.,][]{Vlemmings2024}, but convection does not by itself produce the coherent, sustained outward mass flux required by our solutions. This interpretation is consistent with the absence of spontaneous wind formation in simulations of AGB stars and red supergiants that include a full EoS and radiation transfer \citep[e.g.,][]{Freytag2017,Ma2025}.

Binary orbital decay provides a natural source of pre-existing outward motion in common-envelope events and stellar mergers \citep[e.g.,][]{Ivanova_2013}. If orbital energy deposition produces coherent outward velocities of order $10$--$50\,{\rm km\,s^{-1}}$ at sufficiently high densities, recombination can supply additional energy that unbinds the gas. The corresponding candidate solutions have mass-loss rates of $\sim0.1$--$10\,{\rm M_\odot\,yr^{-1}}$ (Fig.~\ref{fig:param_space}), implying a powerful but short-lived outflow. Although our model assumes spherical symmetry, it is plausible that the outflow is highly aspherical in reality.


Interestingly, our candidate recombination-driven wind solutions occur only for enclosed masses at the critical point $M_{r,c}\lesssim10\,M_\odot$ (Fig.~\ref{fig:param_space}). Given the steady, spherical nature of our models, this result should not be interpreted as a sharp upper limit on the total stellar mass. It may nevertheless indicate that recombination-assisted ejection is favored in systems with lower stellar mass.

Our treatment of radiative transport introduces two principal caveats. First, radiative diffusion is estimated \textit{a posteriori} using LTE Rosseland mean opacities rather than evolved self-consistently with the wind. The quantitative boundary between trapped and diffusive solutions is therefore approximate. Nevertheless, moderately relaxing the trapping criterion leaves the allowed region small, and solutions with $F_{\rm adv}/F_{\rm diff}\sim0.1$ at $v=v_{\rm esc}$ still require large inner-boundary velocities. Our conclusion that recombination does not launch a wind from a hydrostatic envelope is therefore insensitive to moderate uncertainty in the diffusion estimate.

Second, our flux estimate assumes that recombination radiation is locally thermalized. The first generation of recombination photons instead has energies set by atomic bound-free and line transitions rather than by $kT$. Over the densities and temperatures relevant here, however, TOPS frequency-dependent opacity tables indicate that the absorption opacity at characteristic recombination-photon frequencies is generally larger than the Rosseland mean opacity, $\kappa_{\nu,{\rm abs}}\gtrsim\kappa_R$ \citep{TOPS}. These photons therefore have shorter mean free paths than implied by $\kappa_R$, supporting the assumption of efficient trapping and thermalization. At locations where $\kappa_{\nu,{\rm abs}}<\kappa_R$, recombination photons would escape more readily, further weakening recombination-driven acceleration. A self-consistent radiation-hydrodynamic treatment would refine these estimates but is unlikely to change our general conclusions.

Taken together, our results favor a picture in which recombination acts as a supplementary energy source rather than a primary wind-launching mechanism. It can substantially accelerate and unbind an outflow generated by binary interaction or another dynamical process, but it is unlikely to produce a steady wind from an initially hydrostatic star.

\begin{acknowledgments}
We thank Tamar Faran, Ondrej Pejcha, Hanpu Liu, Yoonsoo Kim, and Nicholas Rui for valuable conversations, and Natasha Ivanova, Morgan Macleod,  Brian Metzger, and Ondrej Pejcha for valuable comments on an initial draft.  EQ thanks Stephen Ro for early work together on this problem. This work benefited from interactions supported by the Gordon and Betty Moore Foundation through grants GBMF5076 and GBMF5076.01.
\end{acknowledgments}

\newpage
\appendix 
\section{Velocity gradient at the critical point}\label{app:dvdr}
The velocity gradient in Equation~\eqref{eq:dvdr} can be written as
\begin{equation}
    \frac{d\ln v}{d\ln r} = \frac{{2c_s^2}-\frac{GM_r}{r}}{v^2-c_s^2} \equiv \frac{N}{D}.
\end{equation}
At the critical point, both the numerator and denominator vanish (i.e., $N=D=0$). The velocity gradient must therefore be evaluated using L'H\^{o}pital's rule,
\begin{equation}
    \At{\frac{d\ln v}{d\ln r}}_{r_c} = \At{\frac{dN/dr}{dD/dr}}_{r_c} = \At{\frac{N'}{D'}}_{r_c}, \label{eq:l'hopital}
\end{equation}
where
\begin{align}
    &N' = 2\frac{d c_s^2}{dr} + \frac{GM_r}{r^2} - 4\pi G \rho r, \label{eq:N'} \\
    &D' = 2v^2\frac{d\ln v}{dr} - \frac{d c_s^2}{dr}. \label{eq:D'}
\end{align}

To evaluate $dc_s^2/dr$, we apply the chain rule,
\begin{equation}
    \frac{d c_s^2}{dr} = \At{\frac{d c_s^2}{d\ln\rho}}_T \cdot\frac{d\ln\rho}{dr} + \At{\frac{dc_s^2}{d\ln T}}_\rho \cdot \frac{d\ln T}{dr}.
\end{equation}
Using the adiabatic relation
$d\ln T = (\gamma_3-1)d\ln\rho$, this becomes
\begin{equation}
    \frac{d c_s^2}{dr}
    = c_s^2 \frac{d\ln\rho}{dr} \sqb{\At{\frac{d\ln c_s^2}{d\ln\rho}}_T + (\gamma_3-1) \At{\frac{d\ln c_s^2}{d\ln T}}_\rho}.
\end{equation}
Defining 
\begin{equation}
    A \equiv \At{\frac{d\ln c_s^2}{d\ln\rho}}_T + (\gamma_3-1) \At{\frac{d\ln c_s^2}{d\ln T}}_\rho,
\end{equation}
we obtain
\begin{equation}
    \frac{d c_s^2}{dr} = c_s^2 A \frac{d\ln\rho}{dr}
    = \frac{c_s^2}{r} A \sqb{-\frac{d\ln v}{d\ln r} - 2}, \label{eq:dcs2}
\end{equation}
where the last equivalence uses mass conservation, $4\pi r^2\rho v = \rm constant$.

Substituting Equation~\eqref{eq:dcs2} and the critical-point condition in Equation~\eqref{eq:crit_condition} into Equation~\eqref{eq:N'} yields
\begin{equation}
    N'|_{r_c} = 2\frac{c_s^2}{r_c} A_c\sqb{-\frac{d\ln v}{d\ln r} - 2}_{r_c} + 2\frac{c_s^2}{r_c} - 4\pi G \rho_c r_c. \label{eq:N'c}
\end{equation}
Similarly,
\begin{equation}
\begin{aligned}
    D'|_{r_c} &~= 2\frac{c_s^2}{r} \At{\frac{d\ln v}{d\ln r}}_{r_c} - \frac{c_s^2}{r} A_c\sqb{-\frac{d\ln v}{d\ln r} - 2}_{r_c} \\
    &~= \frac{c_s^2}{r_c} \sqb{(A_c+2) \At{\frac{d\ln v}{d\ln r}}_{r_c} + 2A_c}. \label{eq:D'c}
\end{aligned} 
\end{equation}

We now define $\chi \equiv \At{\frac{d\ln v}{d\ln r}}_{r_c}$. Substituting Equations~\eqref{eq:N'c} and \eqref{eq:D'c} into Equation~\eqref{eq:l'hopital} gives
\begin{equation}
    \chi = \frac{2- 2A_c\chi - 4A_c - B_c}{(A_c+2)\chi + 2A_c},
\end{equation}
where
\begin{equation}
    B_c \equiv \frac{4\pi G\rho_c r_c^2}{c_s^2} = \frac{8\pi \rho_c r_c^3}{M_{r,c}}.
\end{equation}
In practice, we find that $B_c \ll 1$ for most of our wind solutions (this does not imply that the self-gravity of the outflowing mass is negligible everywhere, only that it is small near the critical point; see, e.g., the enclosed mass profiles in Figs \ref{fig:adiabatic_case}-\ref{fig:survived_case}).
Solving the resulting quadratic equation yields
\begin{equation}
\begin{aligned}
    \chi &~ = \frac{-2A_c \pm \sqrt{4-A_c(6+B_c)-2B_c}}{A_c + 2} \\
    &~ \approx \frac{-2A_c \pm \sqrt{4-6A_c}}{A_c + 2}.
\end{aligned}
\end{equation}
where the latter equality is for $B_c \ll 1$.   
A transonic solution requires $\chi > 0$, which implies
\begin{equation}
    \frac{1}{2} > A_c > -2.
\end{equation}

\end{CJK*}

\bibliographystyle{aasjournal}
\bibliography{Bib}

\end{document}